\documentstyle[psfig]{mn2e} 

\title{    The baryon content of the Universe  }

\author[M. Persic \& P. Salucci]{Massimo  Persic $^1$ and Paolo Salucci$^2$ \\ 
       $^1$ Observatory Astronomic, I- 34100 Trieste, Italy\\
       $^2$ SISSA/ISAS, via Beirut 4, I-34014 Trieste, Italy \\
E-mail: {\tt persic@ts.astro.it, salucci@sissa.it}
}

\date{Published, MN, 1992, 258, 14p}

\pubyear{Mon. Not. R. astr. Soc. (1992), 258, 14p}
\begin{document}
\maketitle
\title{ The baryon content of the Universe }
%
%
\begin{abstract}

We estimate the baryon mass density of the Universe due to the stars in galaxies and the hot gas in clusters and groups of
galaxies. The galaxy contribution is computed by using the Efstathiou, Ellis \& Peterson luminosity function, together with
van der Marel and Persic \& Salucci's mass-to-light versus luminosity relationships. We find $\Omega^{stars}_b \simeq 0.002$. For
clusters and groups we use the Edge et al. X -ray luminosity function, and Edge \& Stewart and Kriss, Cioffi \& Canizares'
(gas mass )-luminosity relations. We find $\Omega_b^{gas} \simeq 0.001$. The total amount of visible baryons is then
$\Omega_b \simeq  0.003$, i.e. less than 10 per cent of the lower limit predicted by standard primordial nucleosynthesis,
implying that the great majority of baryons in the Universe are unseen.

\end{abstract}

\begin{keywords}
 Key words: galaxies: fundamental parameters - intergalactic medium -cosmology: observations - dark matter.

\end{keywords}

\section{Introduction}

 The estimation of the baryon mass density of the Universe essentially involves, for each class of objects having a visible
baryon content, an integration over luminosity of the product of the luminosity function (LF), $\phi(L)$, the luminosity,
$L$, and the mass-to-light ratio for the baryon component, $M_b/L$, according to the expression
$$
\rho_b=\sum_T \int  \phi (L) L \Big( {M_b\over L} \Big) dL,
\eqno(1)
$$
where T represents E/S0 galaxies, spiral galaxies, clusters and superclusters, and  $\phi (L)$ and $M_b/L$  refer to the
relevant class of objects.
In the case of galaxies, equation (1) deserves some comments. First, let us note that the baryon mass-to-light ratio
corresponds to the mass-to-light ratio of stars, and not to the dynamical one which usually includes dark matter (DM). In
fact, the presence of DM in the internal parts of spiral galaxies is very evident (Persic \& Salucci 1991), so we cannot
ignore its contribution to the mass of a galaxy. In addition, the importance of DM is a function of luminosity:
low-luminosity galaxies are affected more strongly than high luminosity ones (Persic \& Salucci 1988, 1990). Therefore, it is
essential that the mass-to-light ratio of the visible baryon content be the quantity that enters equation (1). Furthermore,
we emphasize that different populations are present in galaxies of different Hubble types. These have different LFs,
different mean $M_b/ L$ ratios, and different scaling properties
of $(M_b/L)$  with luminosity. Finally, note that, since $\rho_b$  scales as $ h_{50} ^2$ (see equation 1), when
estimated by dynamical arguments the baryon density parameter $\Omega_b=\rho_ b/\rho_c ${ with $\rho_ c = 5 \ h^2_{50} \times
10^{-30}  \  g \ cm^3$ , the critical density } is independent of $h_{50}$.

In practice, the detailed information required by equation (1) has not previously been available. Standard estimates of
$\rho_b$  have assumed a typical value for the visible mass-to-light ratio of galaxies, usually inferred from the observed
dynamics and supposed to be representative for galaxies of all luminosities and Hubble types. Then equation (1) reduces to
$$
 \rho_b= {\cal L}  \Big< \Big({M_b\over L} \Big) \Big>
\eqno(2)
$$

where $\cal L $  is the galaxy luminosity density obtained by integrating the galaxy LF over luminosity and  $<M_b/L> $  is an
assumed mass-to-light ratio. In addition, the hot gas in clusters and groups of galaxies has often been neglected in the
computation of $\Omega_b$ \footnote {Throughout the paper, $h_{50}=H_0/(50 km s^{-1}  Mpc ^{-1}$), where $H_0$ is the Hubble constant.}
 \begin{table*}
\caption{ Notes.$^{(a)}$ Adopted luminosity density, in $10^8 h_{50} L_\odot Mpc^{-3} $  (from $^{[1]}$ Peebles 1971  $^{[2]}$Shapiro 1971; $^{[3]}$ Kirshner, Oemler \& Schechter 1979;  $^{[4]}$Schechter 1976; $^{[5]}$ unspecified; 16J $^{[6]}$Efstathiou et al. 1988;  $^{[7]}$unspecified).$^{(b)}$Adopted mass-to-light ratio, in solar units.$^{(c)}$ Type of adopted mass-to-light ratio, classified according to its procedure of determination.}
\begin{tabular}{l l l l  l}
\hline
\hline
Author                   &  ${\cal L}^{(a)}$        & $<M/L>^{(b)} $      & Type$^{(c)}$  & $\Omega_b $  \\ 
\hline 
Peebles 1971                 & 1.5$^{[1]}$            & $ 10 \ h_{50}$             & dynamical      & 0.02        \\
Gott et al. 1974           &      0.5$^{[2]}$     &     $(2.5-7) \ h_{50}$             &       dynamical   &  $   \geq 0.001$    \\
Olive et al. 1981      &        2 $^{[3]}$           &       $(4-10) \ h_{50}$             & dynamical       &     0.003-0.007        \\
 Boerner 1988          &      0.5$^{[4]}$            &    $(7-10) \ h_{50}  $            & 4.8  dynamical        &  0.011-0.016      \\
Hogan 1990            &       1.5$^{[5]}$          &       2           &    unspecified   &        $0.008  h_{50}^{-1}$ \\
 White 1990            &       1$^{[6]}$          & 1      5            &    stellar   &  $   0.003  h_{50}^{-1}$   \\
Kolb \& Turner 1990  &     1.2 $^{[7]}$           &  $\leq  5 \ h_{50}$           & dynamical        &  $\leq 0.01$      \\  
 
\hline
\label{Summary of baryon density determinations}
\end{tabular}
\end{table*}

For purposes of illustration, in Table l we summarize previous calculations of $\Omega_b$  according to equation (2). Note
that in most cases the {\it dynamical}  rather than the {\it stellar}  mass to-light ratio is used. The range in the adopted values found
in the literature does not reflect observational uncertainties but, rather, real differences in stellar populations, in
proportions of DM and in the reference radius where the  dynamical  mass-to-light ratio is considered.
Detailed information has recently been published on the properties of stellar mass-to-light versus luminosity relations for
E/S0 and S galaxies (van der Marel 1991; Persic \&
Salucci 1991), as well as on the relation between hot gas 
content and X-ray luminosity for clusters of galaxies (Edge \& Stewart 1991a, b). Also, LFs have become available for E/S0
and S galaxies separately (Efstathiou, Ellis \& Peterson 1988) as well as for clusters (Edge et al. 1990). It is therefore
possible to obtain dynamical measures of the visible mass associated with these structures. The aim of this paper is to use
these advances in estimating the value of $\Omega_b$. The value we find,  $\Omega_b \simeq 0.003$ is lower than some
previous estimates. Compared to current standard nucleosynthesis predictions,
our estimate aggravates the problem of the 'missing baryons'.

\section{THE MEAN BARYON DENSITY FROM GALAXIES AND CLUSTERS}

In all cases, the LF will have the usual Schechter (1976) form:
 $$
\phi(L) {dL\over L_*}=  \phi_* \Big({L\over L_*}\Big)^{-\alpha} {dL\over L_*}
\eqno(3)
$$
where $\phi_*$  is a normalization constant, $\alpha$  is the slope of the LF at low luminosities, and $L_* $ is the
luminosity corresponding to the 'knee' of the LF; and the baryon mass-to-light versus luminosity relation will be a power
law:
 $$
{M_b\over L}=a \Big({L\over L_*} \Big)^\eta
\eqno(4)
$$
with $A$ the mass-to-light ratio at the characteristic luminosity $L_*$. Inserting equations (3) and (4) into equation (2) we
get
$$
\rho_b =\phi_* L_* A \int^{x_{max}}_{x_{min}} x ^{1-\alpha+\eta} e^{-x} dx 
\eqno(5) 
$$
 
where $x= L/L_*$ and $x_{min}$ and $x_{max}$ represent the observed minimum and maximum luminosities for a given class of
objects. The values of the parameters $\phi_*$, $\alpha$, $L_*$, $x_{min}$,   $ x_{max}$, $A$ and $T$  are observationally known and will be
chosen accordingly for each class of objects.
We shall treat separately elliptical (and S0) galaxies, spiral galaxies, groups and clusters of galaxies.

\subsection{Ellipticals} 

From Efstathiou et al.'s (1988) field-galaxy LF we get $\phi_* =8.5 \times 10^{-4} h_{50}^3 Mpc^ {-3}$, $\alpha =0.48 $ and
$L_*=3.3 \times 10^{10} h_{50}^{-2}  L_\odot $;  we take $x_{min} = 0.04$  and $ x_{max} = 8$. There is strong evidence that
ellipticals do not have DM inside their optical radii, so their dynamical masses at the optical radius and their stellar
masses coincide (see Djorgovski \& Davis 1987). From van der Marel (1991) we have $A =4 h_{50}  M_\odot/L_\odot $  and $\eta =
0.35$  (all values refer to the B band). Inserting these values into equation (5), the integral takes a value of 0.97, and we
then get $ \rho_b= 7.4 \times 10^{-33} h^2_{50} \ g \  cm^{-3}$.  Thus the baryon contribution of elliptical galaxies to the mean density is 

$$
\Omega^{\rm (E + S0)}_b= 1.5 \times 10^{-3}  \ \ \  \  {\rm  (E + S0 \  galaxies).}
\eqno(6) 
$$

\subsection{Spirals}

From Efstathiou et al. (1988) we get $\phi_* =1.1 \times  10^{-3} h_{50}^3\ Mpc^ {-3} $,  $\alpha =1.24  $ and $L_*=4.2 \times
10^{10} h_{50}^-{2}  L_\odot $;  we take $x_{min} = 0.01 $ and $ x_{max} = 8$.  Dark and visible matter are already well mixed
in the optical regions of spirals (e.g. Persic \& Salucci 1991). From Persic \& Salucci (1990) dynamical disk/halo mass
decomposition of rotation curves, we get  $A =1.2 h_{50}  M_\odot/L_\odot $  and $\eta  = 0.35$  (all  values refer to the B
band). Inserting these values  into equation (5), the integral takes a value of 0.94, so we get $ \rho_b= 3.5 \times
10^{-33} h^2_{50}\ g \ cm^{-3}.$  corresponding to

 $$
\Omega^{\rm (S)}_b \simeq  0.7  \times 10^{-3}  \ \ \ \    {\rm  (S \ galaxies)}.
\eqno(7)
$$

\subsection{Rich clusters}

The hot, X-ray emitting, diffuse intracluster gas is a distinctive prominent baryon component of clusters (e.g. Sarazin
1986). From Edge et al. (1990) we get   $\phi_* =6.9 \times  10^{-8} \  h_{50}^3 \ Mpc^ {-3} $,  $\alpha =1.65  $ and $L_*=8.1 \times 10^{44}\  h_{50}^{-2} erg/sec $;  we
take $x_{min} = 0.012$  and $ x_{max} = 2.5$; relative to the gas mass within $0.5 h_{50}^{-1}$ Mpc radius, from Edge \& Stewart (1991a) we take  
 $A =50  \ h_{50}^{-0.5}  g\  erg^{-1} \ sec  $  and $\eta  = -0.62$ 
 (AlI the X-ray data are relative to the (2 - 10) keV band.) Inserting the above values into equation (5),
the integral takes a value of 7.6,  corresponding to (a contribution to) $\Omega_b=1.4
\times  10^{-4}$. This estimate refers to the gas mass contained within $0.5 h_{50}^{-1} \ Mpc$ radius (i.e., within - 2 optical core radii, see Bahcall 1977).
In, at least some cases, however, there is evidence (Jones \& Forman 1984) for significant amounts of  gas at larger
distances, $\sim 3 h_{50}^{-1}  Mpc$. This distance is a good measure of the typical virialization radius of  rich clusters
[e.g., from a recent study of  a large rich-cluster sample we have $R_{vir}=(3.0 \pm 0.2) h_{50}^{-1} Mpc$ (Biviano et al. 1992)], and
corresponds to the Abell (1958) radius, ie. the radius roughly encompassing most of  the cluster's member galaxies. The
hot-gas surface-brightness profile can be described in terms of  the so-called $\beta$-model,
$$
S(r) =S_0 \ [1 + (r/a)^2]^{-3\beta+1/2} 
$$
with $a$ (the core radius) and $\beta$ as free parameters, which implies a de-projected spatial gas density distribution of
the form
$$
\rho_{gas} (r) =  \rho_{0} [1+ (r/a)^2]^{-3\beta/2}
$$
(e.g., Henriksen \& Mushatzky 1985; David et al. 1990). Fits to Einstein IPC data far a survey of clusters give for $\beta$
an average value of  $<\beta> = 0.66 \pm  0.10$ (see Jones \& Forman
1984; David et al 1990). Therefore, on average, beyond a few core radii the hot- gas density falls off as $\rho_{gas} \propto r^{-2}$, and
the mass in gas rises linearly with radius. We therefore estimate that the hot-gas contribution of rich clusters from within
the Abell radius is
 	 
 $$
\Omega^{\rm (cl. gas)}_b \simeq  8.6  \times 10^{-4} h_{50}^{-1.5}  \ \ \ \  {\rm    (intracluster  \ gas)}
\eqno(8)
$$	 
In addition to the diffuse gas, further baryons in clusters are contributed by the stellar component  of  the member
galaxies. However, recent direct estimates (see Edge \& Stewart 1991b) indicate that, in the cores of clusters, the stellar
mass is up to  a factor of  3 smaller than the gas mass.  Further out   there are indications that the galaxy distribution
drops more rapidly than the gas (e.g., Eyles et al. 1991), so  that within a few core radii the stellar mass is
significantly lower than the gas mass (roughly 10 per cent for a stellar mass-to-light ratio of 4; see David et al. 1990),
and at the Abell radius the stellar component is negligible compared to  the gas mass.

\subsection {Poor clusters and groups}

A population of  poor clusters and groups of  galaxies must certainly contribute to the baryon content. The information
on the LF and the gas content at these mass scales is quite scanty. We extrapolate the Edge et al. (1990) c1uster X-ray LF
through the range of luminosities typical far these structures, $41 \leq  log L(2-10 keV) \leq 43 $ (see Kriss, Cioffi \& Canizares
1983; Bahcall, Harris \& Rood 1984; see Bahcall 1979 for a discussion of  the continuity of  cluster and group LFs). Thus in
equation (5) we take the parameters of  the Edge et al. (1990) cluster LF (see above), with $x_{min}=1.2 \times 10^{-4}$  and $x_{max}=0.012$.
Relative to the hot-gas content within $0.5  h_{50}^{-1} Mpc$ radius, the {\it Einstein}  data of  Kriss et al. (1983) imply $A =65 h_{50}^{-0.5}$   and $\eta  = - 0.5$  (in the 0.5-4.5 keV band). The integral in equation (5) takes a value of  13, so we derive   $\rho_b = 1.6 \times  10^{-33} g \ cm^{-3} \ h_{50}^{0.5}$  
 corresponding to a contribution to   $\Omega_b  \simeq  3.2  \times 10^{-4} h_{50}^{-1.5}   $      Allowing far any hot gas extending
out to a typical group virialization radius (  $\sim  1 h_{50}^{-1} Mpc$, see Pisani 1990), by the same argument used for the rich clusters
we get
 $$
\Omega^{\rm (gr. gas)}_b \simeq  6.4   \times 10^{-4} h_{50}^{-1.5}  \ \ \ \    {\rm  (intragroup \ gas).}
\eqno (9)
$$
 
\section{DISCUSSION}

From   equations (6)-(9), we conclude that  the combined baryon contribution of  galaxies and clusters to  the mean density
is:
$$
\Omega_b \simeq  2.2   \times 10^{-3} +   1.5 \times 10^{-3} h_{50}^{-1.5 }
\eqno (10)
$$

As $1\leq h_{50} \leq 2 $  with a currently favored value of  = 1.5 (e.g., Pierce \& Tully 1988), the cluster/group density contribution(see  equations 8 and 9) is probably $\Omega_{gas} \simeq 0.001.$. The
estimated contribution of  the stellar populations of  galaxies (see equations 6 and 7) is $\Omega_* \simeq 0.002.$ This value is in
very good agreement with  the cosmological mass density of  the damped Ly$\alpha$ system, which are un likely to be proto-galactic 
discs and would give a present-day contribution of $ \Omega_b(Ly\alpha) \sim  0.002 $(see Wolfe 1988).   
Note that the explicit inclusion of  the general trend of  decreasing stellar mass-to-light ratio with  decreasing
luminosity and the accounting for different morphological classes have the effect of reducing the estimated baryon
contribution of galaxies. In fact, the often-assumed values of  $M_b/L_B = 8  h_{50}$   for ellipticals and $M_b/L_B = 3  h_{50}$ for spirals
actually refer only to the brightest objects (i.e. $L_B> 6 L_*$). Assuming such values as typical would overestimate the baryon
contribution from stars by a factor of - 2. As an example (of it), let us take our  mass-to-light ratios at $L_*$ as the typical
values (i.e.  4  far ellipticals and 1.2  far spirals) to be used in equation (2). Let us also use separate LFs to compute
the luminosity densities due to ellipticals and spirals (i.e., $0.25 \times 10^8  L_{B_\odot} h_{50}^{-2}  Mpc^{-3}$ and $ 0.54 \times 10^8  L_{B_\odot} h_{50}^{-2}   Mpc^{-3}$ respectively, from
Efstathiou et al. 1988). Then equation (2) would be written as
$$
\rho_b=[{\cal L }<(M_b/L_B)> ]_{E/S0} +[{\cal L} <(M_b/L_B)> ]_S 
$$
$$
=([0.25 \times  10^8 \times  4]+[0.54 \times  10^8   \times 1.2])  h_{50}^{-2}   M_\odot /Mpc^{-3}
$$
Thus we would find (from the above equation)  contributions to $\Omega_{b,galaxies}$ of  $1.4 \times 10^{-3}$ and $0.9 \times 10^{-3}$ for the two classes separately. This example shows that 
the discrepancy between some previous estimates of $\Omega_*$ (claimed to agree with nucleosynthesis limits) and our own estimate
arises mainly because these calculations used {\it dynamical} and unrealistic mass-to-light ratios and, to a lesser extent,
because they did not use  separate LFs for different morphologies.
Further baryons, in the farm of  hot ($T \geq 10^7 K$) diffuse gas, could be supplied by superclusters. However, the observational
limit to any diffuse X-ray emission from candidate supercluster cores (see Persic et al. 1990), alongside the spatial
frequency of  superclusters (ie., $\sim 10^8 h_{50}^{3} Mpc^{-3}$), indicates a negligible contribution:
$$
\Omega^{\rm SC, gas}_b \sim 10^{-5}
$$

Another reservoir of  baryons could be cold diffuse HI gas in the local intergalactic medium (IGM). However, the absence
of  the Gunn-Peterson (1965) trough on local scales
places a severe upper bound to any such contribution: $\Omega_{HI} < 5 \times 10^{-7}$ (See Davidsen et al 1977)

We therefore expect that our derived value of $\Omega_b$ in equation (10) gives a close representation of the actual visible baryon
content of the local Universe. [Fabian's (1991) recent estimate of the baryon density in the Shapley supercluster, $ \Omega_b \simeq     0.18\  h_{50}^{-1.5 }$ is not necessarily in contradiction with our estimate. In fact, Fabian's high value relates to a most exceptional
region of $ \sim 35 \   h_{50}^{-1} Mpc$ size and with $\delta \rho/\rho \simeq 0.8 \Omega_0^{-1}$, while our low value refers to the whole nearby Universe out to a
radius of $ \sim 300 h_{50}^{-1} \ Mpc$ and with $\delta \rho/\rho \sim 0$] Our value, $\Omega_b \simeq 0.003$, is substantially lower than the prediction of standard
cosmic nucleosynthesis, $0.04< \Omega_b h_{50}^2$ with a probable value of $\Omega_b \simeq 0.06$  (e.g., Kolb \& Turner 1990; Peebles et al. 1991).
Comparing the observed and the predicted baryon abundances, we conclude that {\it the stars and gas
of galaxies and clusters/groups account for only $\leq 10$  per cent of the primordially synthesized baryons} (see also Hogan
1990; Kolb \& Turner 1990).
Where are the 90 per cent of missing baryons? We have
only taken an inventory of the visible baryons associated with visible structures. Additional baryons, unaccounted for by
the present census, can either be clumped in some dark form, for instance forming dark haloes around galaxies, or be
distributed in a diffuse ionized background. [Note that the diffuse DM associated with clusters, exceeding the gas mass by
at most a factor of - 3 (e.g., Eyles et al. 1991), could not solve the discrepancy with standard nucleosynthesis even if it
were completely baryonic.] Let us consider each of the two possibilities in turn.

{\it (i) Baryons in haloes.} The dark haloes of spiral galaxies may well extend out to 10-20 times the size of the optical discs.
Thus it is not difficult to conceive that, by integrating the dynamical mass-to-light ratio of galaxies (computed at such
extended radii) over the LP, the nucleosynthesis value of $\Omega_b \simeq  0.06$ might be easily reached. An attractive way of
hiding the missing baryons, therefore, is to assume that they constitute the DM in galaxy haloes. This possibility may find
support  from the evidence that cooling flows may be producing baryonic DM in the form of low-mass stars or brown dwarfs
(e.g., Fabian, Nulsen \& Canizares 1984). Based on cooling flow analogies, Thomas \& Fabian (1990) have argued that baryonic
DM only forms on fairly large mass-scales where gas cooling is quasi-static, while Ashman (1990) has suggested that baryonic
DM forms on galactic and subgalactic scales, following rapid gas cooling (Ashman \& Carr 1991).

{\it (ii) ionized IGM}. The second possibility is to suppose that galaxy formation is extremely inefficient, so that only 10 per
cent of gas in the Universe is now in collapsed structures such as galaxies. This could arise if gas was never incorporated
into galaxies, or if gas was expelled from protogalaxies by supernova explosions or galactic winds (Bookbinder et al.
1980). In this scenario, most baryons in the Universe now constitute a smooth ionized IGM, in agreement with the lack of any
Gunn-Peterson trough in the spectra of quasars both at high redshifts (see Sargent \& Steidel 1990) and locally (Davidsen et
al. 1977). There is no shortage of proposed
sources of ionization. Among conventional ad hoc sources there are Population III stars (Carr 1990), dwarf and starbursts
galaxies (Silk, Wyse \& Shields 1987; Songaila, Cowie \& Lilly 1990) and obscured quasars (Miralda-Escude' \&
Ostriker 1990). Alternatively, a non-conventional self consistent explanation involves the radiative decay of cosmological
dark matter  particles (Rephaeli \& Szalay 1981; Sciama 1982).
We conclude by stressing that the result presented in  this paper, in connection with the dynamical estimates of the total
mass density of the Universe on scales  $\sim 10 h_{50}^{-1} Mpc$, $\Omega_0 \geq 0.2$ (see Davis et al. 1980; Davis \& Peebles 1983) supports the
conventional view on the need for non-baryonic extra dark matter.

\section{ACKNOWLEDGMENTS}
We thank Dennis Sciama for stimulating conversions that motivated us to undertake this calculation and for continuous
encouragement. We also thank Keith Ashman for discussing with us the role of baryonic dark matter. Finally, we acknowledge
the referee's useful advice which has helped us to improve the presentation of this work.

\section{References}

Abell, G. O., 1958. Astrophys. J. Suppl, 3,211. \\
Ashman, K. M., 1990. Astrophys. J., 359, 15.  \\
Ashman, K. M. \& Carr, B. J., 1991. Mon. Not. R. Astr.Soc., 249, 13.   
Bahcall, N. E., 1977. Ann. Rev. Astr. Astrophys., 15,505.\\
Bahcall, N. E., 1979. Astrophys. J., 232,689.\\
Bahcall, N. E., Harris, D. E. \& Rood, H. J., 1984. Astrophys. J. Lett.,
284, L29.\\
Biviano, A., Girardi, M., Giuricin, G., Mardirossian, F. \& Mezzetti,
M., 1992. Astrophys. J., in press. \\
Bookbinder, J., Cowie, L. L., Krolik, J. H., Ostriker, J. P. \& Rees, M.,
1980. Astrophys. J., 237,647. \\
Borner, G., 1988. The Early Universe, Springer-Verlag, Berlin.  \\
Carr, B.J., 1990. Comm. Astrophys., 14,257.  \\
David, L.P., Arnaud, K. A., Forman, W. \& Jones, c., 1990. Astrophys. J., 356,
32. \\
Davidsen, A. F., Hartig, G. F. \& Fastie, W. O., 1977. Nature, 269. 203. \\
Davis, M. \& Peebles, P. J. E., 1983. Astrophys. J., 267, 465.  \\
Davis, M., Tonry, J., Huchra, J. \& Latham, D. W., 1980. Astrophys. J. Lett.,
238, L113.  \\
Djorgovski, S. \& Davis, M., 1987. Astrophys. J., 313, 59.  \\
Edge, A. C. \& Stewart, G. C., 1991a. Mon. Not. R. astr. Soc., 252, 414. \\
Edge, A. C. \& Stewart, G. C., 1991b. Mon. Noi. R. aslr. Soc., 252, 428. \\
Edge, A. c., Stewart, G. c., Fabian, A. C. \& Arnaud, K. A., 1990.
Mon. Not. R. astr. Soc., 245,559.\\
Efstathiou, G., Ellis, R. S. \& Peterson, B. A., 1988. Mon. Not. R.
astr. Soc., 232,431.\\
Eyles, C. J. et al., 1991. Astrophys. J., 376,23.\\
Fabian, A. C., 1991. Mon. Not. R. astr. Soc., 253, 91p.\\
Fabian, A. c., Nulsen, P. E. J. \& Canizares, C. R., 1984. Nature,
311,733.\\
Gott, J. R., Gunn, J. E., Schramm, D. N. \& Tinsley, B. M., 1974,
Astrophys. J., 194, 543.\\
Gunn, J. E. \& Peterson, B. A., 1965. Astrophys. J., 142, 1633. \\
Henriksen, M. J. \& Mushotzky, R F., 1985. Astrophys. J., 292, 441. \\
Hogan, C. J., 1990. In: Baryonic Dark Matter, p. 1, eds Lynden-Bell,
D. \& Gilmore, G., Kluwer, Dordrecht.\\
Jones, C. \& Forman, w., 1984. Astrophys. J., 276, 38. \\
Kirshner, R. P., Oemler, A., Jr \& Schechter, P. L., 1979. Astr. J., 84, 951.\\
Kolb, E. W. \& Turner, M. S., 1990. In: The Early Universe, Addison­Wesley 
Publishing Company, California.\\
Kriss, G. A., Cioffi, D. F. \& Canizares, C. R., 1983. Astrophys. J., 272, 439. \\
Miralda-Escude, J. \& Ostriker, J. P., 1990. Astrophys. J., 350, 1. \\
Olive, K. A., Schramm, D. N., Steigman, G., Turner, M. S. \& Yang,J., 
1981. Astrophys. J., 246,557.\\
Peebles, P. J. E., 1971. Observational Cosmology, Princeton University Press, 
Princeton, NJ.\\
Peebles, P. J. E., Schramm, D. N., Turner, E. L. \& Kron, R. G., 1991.
Nature, 352,769.\\
Persic, M. \& Salucci, P., 1988. Mon. Not. R. astr. Soc., 234,131. \\
Persic, M. \& Salucci, P., 1990. Mon. Not. R. astr. Soc., 245,577. \\
Persic, M. \& Salucci, P., 1991. Astrophys. J., 368,60.\\
Persic, M., Jahoda, K., Rephaeli, Y., Boldt, E., Marshall, F. E.,
Mushotzky, R. F. \& Rawley, G., 1990. Astrophys. J., 364, 1.\\
Pierce, M. J. \& Tully, R. B., 1988. Astrophys. J., 330,579.\\
Pisani, A., 1990. PhD thesis, SISSA, Trieste.\\
Rephaeli, Y. \& Szalay, A. S., 1981. Phys. Lett., 106B, 73.\\
Sarazin, C. L., 1986. Rev. Mod. Phys., 58, 1.\\
Sargent, W. L. W. \& Steidel, C. C., 1990. In: Baryonic Dark Matter,
p. 223, eds Lynden-Bell, D. \& Gilmore, G., Kluwer, Dordrecht. \\
Schechter, P. L., 1976. Astrophys. J., 203, 297.\\
Sciama, D. w., 1982. Mon. Not. R. astro Soc., 198, 1p.\\
Shapiro, S., 1971. Astr. J., 76,291.\\
Silk, J., Wyse, R. \& Shields, G. H., 1987. Astrophys. J. Lett., 322, L59.\\
Songaila, A., Cowie, L. L. \& Lilly, S. J., 1990. Astrophys. J., 348, 371.\\
Thomas, P. A. \& Fabian, A. c., 1990. Mon. Not. R. astr. Soc., 246, 156. \\
van der Marel, R. P., 1991. Mon. Not. R. astr. Soc., 253,710. \\
White, S. D. M., 1990. In: Physics of the Early Universe, Proc. 36th
Scottish Universities Summer School in Physics, eds Peacock,
J. A., Heavens, A. F. \& Davies, A. T.\\
Wolfe, A. M., 1988. In: QS0Absorption Lines: Probing the Universe, 
Proc. Space Telescope Science Institute Symp. No.2, p.297, 
eds Blades, J. C., Turnshek, D. \& Norman, C. A., Cambridge 
University Press, Cambridge.

\end{document}